\documentclass[aps,pra,twocolumn,superscriptaddress,floatfix,
nofootinbib,showpacs,longbibliography]{revtex4-1}
\usepackage[utf8]{inputenc}  
\usepackage[T1]{fontenc}     
\usepackage[british]{babel}  
\usepackage[sc,osf]{mathpazo}
\usepackage[scaled=0.86]{berasans}  
\usepackage[colorlinks=true, citecolor=blue, urlcolor=blue]{hyperref}  
\usepackage{graphicx}
\usepackage[babel]{microtype}  
\usepackage{amsmath,amssymb,amsthm,bm,amsfonts,mathrsfs,bbm} 

\usepackage{xspace}  
\usepackage{pgf,tikz}
\usepackage{xcolor}
\usepackage{multirow}
\usepackage{array}
\usepackage{bigstrut}
\usepackage{braket}
\usepackage{color}
\usepackage{natbib}
\usepackage{multirow}
\usepackage{mathtools}
\usepackage{float}
\usepackage{xcolor,colortbl}
\usepackage{physics}
\usepackage{amsmath}
\usepackage{color}
\usepackage{array} % required for text wrapping in tables
\usepackage[justification=justified, format=plain]{subcaption}
\usepackage[justification=raggedright]{caption}

\newcommand{\be}{\begin{equation}}
\newcommand{\ee}{\end{equation}}
\newcommand{\ba}{\begin{eqnarray}}
\newcommand{\ea}{\end{eqnarray}}

\def\>{\rangle}
\def\<{\langle}

\begin{document}
	
\title{ Efficient detection of nonclassicality using moments of the Wigner function}

\author{Bivas Mallick}
\email{bivasqic@gmail.com}
\affiliation{S. N. Bose National Centre for Basic Sciences, Block JD, Sector III, Salt Lake, Kolkata 700 106, India}

\author{Sudip Chakrabarty}
\email{sudip27042000@gmail.com}
\affiliation{S. N. Bose National Centre for Basic Sciences, Block JD, Sector III, Salt Lake, Kolkata 700 106, India}

\author{Saheli Mukherjee}
\email{mukherjeesaheli95@gmail.com}
\affiliation{S. N. Bose National Centre for Basic Sciences, Block JD, Sector III, Salt Lake, Kolkata 700 106, India}

\author{Ananda G. Maity}
\email{anandamaity289@gmail.com}
\affiliation{Networked Quantum Devices Unit, Okinawa Institute of Science and Technology Graduate University, Onna-son, Okinawa 904-0495, Japan}

\author{A. S. Majumdar}
\email{archan@bose.res.in}
\affiliation{S. N. Bose National Centre for Basic Sciences, Block JD, Sector III, Salt Lake, Kolkata 700 106, India}

%%%%%%%%%%%%%%%%%%%%%%%%%%%%%%%%%%%%%%%%%%%%%%%%%%%
\begin{abstract}
States with a negative Wigner function, a significant subclass of nonclassical states, serve as a valuable resource for various quantum information processing tasks. Here, we provide a criterion for detecting such quantum states characterized by a negative Wigner function. Our method relies on evaluating moments of the Wigner function, which involves computing simple functionals and can be implemented in a real experiment without the need for full state tomography or Wigner function reconstruction. We provide explicit examples to support our detection scheme. Further, we propose an experimental method utilizing the mode SWAP operator to realize these moments in a real experiment.
\end{abstract}
\maketitle

%%%%%%%%%%%%%%%%%%%%%%%%%%%%%%%%%%%%%%%%%%%%%%%%%%%%
\section{Introduction}\label{s1}
 Quantum optical systems \cite{agarwal2012quantum, gerry2023introductory}
involving both discrete variable (DV) and continuous variable (CV) states
\cite{review_braunstein}, play a significant role in the development of future quantum technologies by offering a promising and powerful platform for efficiently engineering large and scalable multimode quantum states. Traditionally in CV systems, the states residing in infinite-dimensional Hilbert spaces can effectively be represented in terms of quasiprobability distribution over phase-space $\{q,p\}$. Among these  quantum states, a particular type of quasiprobability distribution is characterized by Gaussian statistics and can be succinctly described by their first and second statistical moments \cite{review_gaussian,adesso2014continuous}. While these states can exhibit various intriguing quantum phenomena, they can be simulated classically \cite{eisert2012positivewigner} indicating that they are not sufficient on their own for executing quantum information protocols. On the other hand, there exist quantum states, including non-Gaussian states \cite{albarelli2018resource,takagi2018convex,chabaud2020stellar,
adesso2014continuous,chowdhury2013nonlocal}, that offer genuinely non-classical features, providing greater advantages in optical quantum communication, quantum information processing, quantum metrology, quantum computation, and many more tasks \cite{giedke2002characterization,niset2009no,bartlett2002efficient,
adhikari2008teleportation,adhikari2008broadcasting}. \\

The notion of non-classicality of an optical quantum state was proposed by Mandel \citep{mandel1979sub} in terms of the deviation of photon number statistics from a Poisson distribution which is a typical characteristic of classical states (e.g., coherent states) \citep{glauber1963coherent}. Since then, there have been many different approaches to classify non-classical states and their behavior \cite{dodonov2002nonclassical,lee1991measure, rahimi2013quantum}. Typically,  non-classical states are defined as those whose Glauber-Sudarshan P-representation \cite{sudarshan,glauber1963coherent}  fails to meet the criteria of an ordinary probability distribution \cite{tan2020negativity,titulaer1965correlation}. Though non-classicality can arise in states with a positive Wigner function, a significant subclass of non-classical states includes those with a negative Wigner function \cite{wigner1932quantum, kenfack2004negativity,cormick2006classicality}. States with negative Wigner function offer promising advantages in several quantum information processing tasks \cite{giedke2002characterization,niset2009no,bartlett2002efficient,galvao2005discrete}. Moreover, states with positive Wigner functions can be efficiently simulated on classical computers, meaning that Wigner negativity is necessary for any quantum speedup in such systems \cite{eisert2012positivewigner}. Thus, Wigner negativity serves not only as a non-classical characteristic but also as a necessary resource for achieving computational advantages in quantum computing.\\

The states with negative Wigner function (or Wigner negativity for short) offer promising advantages in information processing tasks such as quantum state distillation \cite{giedke2002characterization, fiuravsek2002gaussian,eisert2002distilling}, quantum error-correction\cite{niset2009no}, quantum computing \cite{bartlett2002efficient,eisert2012positivewigner}, quantum computational speedup \cite{galvao2005discrete}, and so on.   Wigner negative states have been experimentally generated in optical cavities \citep{yoshikawa2013creation}, trapped-ion \citep{fluhmann2019encoding}, superconducting circuits \citep{lu2021propagating}, optical and optomechanical experiments with photon-number resolving detectors \cite{innocenti2022nonclassicality}. These findings highlight states possessing Wigner negativity as a powerful resource for implementing various quantum information processing tasks in real systems.\\

Detection of Wigner negative states is a prerequisite for exploring their applications in any legitimate quantum information processing task. Several studies have explored the challenge of detecting Wigner negativity
\cite{chabaud2021witnessing,cimini2020neural}.
Wigner negativity can  be detected by Wigner function reconstruction via direct or indirect methods. The direct method involves obtaining the Wigner function directly at a specific point in phase-space \citep{leibfried1996experimental, smithey1993measurement, banaszek1999direct, lutterbach1997method,  bertet2002direct, winkelmann2022direct}, whereas in the indirect method, the quantum state is reconstructed first, and subsequently, the Wigner function is derived by a mathematical transformation \cite{lv2017reconstruction, lvovsky2001quantum, fluhmann2020direct}. However, obtaining the Wigner function of an unknown quantum state is costly in terms of resources due to the substantial number of measurements required. This process demands a tomographically complete set of measurements, typically involving multiple measurement configurations \cite{smithey1993measurement}. This task becomes more challenging in continuous variable systems because the Hilbert space dimension of these CV quantum states is infinite \cite{lvovsky2009continuous,Rahimi-Keshari_2011}. Other methods for the detection of Wigner negativity exist, such as those based on numerical simulation techniques \citep{cimini2020neural}, or involving construction of witness operators \cite{horodecki1996necessary,terhal2001family,lewenstein2001characterization,mari2011directly,chabaud2021witnessing,kiesel2012universal}. However, the witness operator formalism is state-dependent and its effectiveness of is based on prior knowledge of the state to be detected.\\

Driven by the motivation to utilize Wigner negativity as a resource, along with recent efforts in designing useful tools for its detection, in this work we present a criterion for identifying the signature of Wigner negativity. Our formalism is not restricted to  CV states only but is also valid for  
single- and multi-mode  DV states which have a continuous Wigner function representation. Our approach based on the moment criterion, enables the evaluation of simple functionals, making it experimentally less demanding compared to methods requiring full state tomography or Wigner function reconstruction \cite{lvovsky2009continuous,d2001quantum, d2003quantum}. Importantly, we demonstrate that the first three moments are sufficient for identifying Wigner negativity. We provide explicit examples to support our detection scheme and show that these moments can be efficiently obtained in real experiments using the mode SWAP operator. Our criterion does not require any prior knowledge about the state, in contrast to witness-based detection schemes that are state-dependent. Additionally, moment-based approaches are advantageous in terms of the number of copies required to determine Wigner negativity \cite{ghalaii2017scheme}. The number of measurements needed for tomography generally scales exponentially with the system size for non-Gaussian states, whereas a polynomial number of state copies suffices for moment-based approaches.\\

The paper is organized as follows. In Sec. \ref{s2}, we present an overview of the fundamental mathematical concepts related to continuous variable bosonic systems along with an introduction to the Wigner function and its phase-space representation. This section also revisits the moment criteria, previously proposed for entanglement detection. In Sec. \ref{s3}, we introduce the concept of the moments of the Wigner function and present our detection criterion along with some explicit examples. An experimental proposal for evaluating the moments of the Wigner function using the mode SWAP operator is presented in Sec. \ref{s4}. Finally, in Sec. \ref{s5}, we summarize our main findings along with some future perspectives.

%%%%%%%%%%%%%%%%%%%%%%%%%%%%%%%%%%%%%%%%%%%%%%%%%%

\section{Preliminaries}\label{s2}

A quantum system consisting of $N$ canonical bosonic modes is characterized by a Hilbert space, 
\begin{equation}
    \begin{split}
        \mathcal{H} = \bigotimes_{k=1}^N \mathcal{H}_k \nonumber
    \end{split}
\end{equation}
where each $\mathcal{H}_k$ associated with a single mode denotes a separable, infinite-dimensional Hilbert space. The countable orthonormal basis vectors are the elements of the single mode Fock basis $\{\ket{n}\}_{n \in \mathbb{N}}$. $\mathcal{D(\mathcal{H})}$ represents the set of positive operators with unit trace acting on $\mathcal{H}$ where the density operator of the system resides. Let us denote the annihilation and creation operators in mode $k$ by $\hat{a}_k$ and ${\hat{a}^{\dagger}_k}$ respectively, which satisfy the bosonic commutation relations,
\begin{equation}
    \begin{split}
        [\hat{a}_k, \hat{a}^{\dagger}_l] = \delta_{k l} \mathbb{I}  \hspace{0.3mm},\hspace{3mm} [\hat{a}_k, \hat{a}_l] = [\hat{a}^{\dagger}_k, \hat{a}^{\dagger}_l] = 0. \nonumber
    \end{split}
\end{equation}
Their action on the Fock basis is defined as:
\begin{equation}\nonumber
\begin{split}
&\hat{a} \ket{n} = \sqrt{n} \ket{n-1} \hspace{0.3cm} \text{for} \hspace{0.15cm} n \in {\mathbb{N}}^{*}\hspace{3mm} \text{with} \hspace{3mm} \hat{a} \ket{0} =  0, \hspace{0.3cm} \\&
\hat{{a^{\dagger}}} \ket{n} = \sqrt{n+1} \ket{n+1} \hspace{0.2cm} \text{for} \hspace{0.15cm} n \in {\mathbb{N}}.
\end{split}
\end{equation}
The quadrature phase operators for mode $k$ are defined as,
\begin{equation}
        \hat{x}_k = \frac{(\hat{a}_k + \hat{a}^{\dagger}_k)}{\sqrt{2}},\hspace{3mm}\hat{p}_k = \frac{(\hat{a}_k - \hat{a}^{\dagger}_k)}{\sqrt{2}i}, 
\end{equation}
with $[ \hat{x}_k, \hat{p}_k ] = i \mathbb{I}$. For convenience, we shall use natural units for which $\hbar=1$ throughout the manuscript.

In the next subsection, we provide a brief introduction to the Wigner function. Throughout the manuscript, we shall use the terminologies commonly used in the field of quantum information.
%%%%%%%%%%%%%%%%%%%%%%%%%%%%%%%%%%%%%%%%%%%%%

\subsection{Wigner function}
The Wigner function is a quasiprobability distribution function used in the phase-space formulation of quantum mechanics \cite{cahill1969density,leonhardt1997measuring}. Its ability to exhibit negativity distinguishes its quantum behavior from classical. This function provides a complete description of the quantum state by representing it in a combined position and momentum space. Let ($x_k, p_k$) be the position and momentum quadratures of $k$-th mode, with the corresponding vectors represented by $\vec{x} = (x_1,x_2,...,x_N),~ \vec{p} = (p_1,p_2,...,p_N)$. The Wigner function for an $k$-mode system corresponding to density matrix $\hat{\rho}$ is defined as \cite{wigner_function_for_pedestrians}
\begin{equation}
W(\vec{x}, \vec{p}) = \frac{1}{{(2 \pi) }^k} \int\limits_{-\infty}^{\infty} \bra{ \vec{x} +\frac{\vec{y}}{2}}  \hat{\rho} \ket{\vec{x} - \frac{\vec{y}}{2} } e^{-i \vec{p} \cdot \vec{y}} \hspace{0.1cm}d^k y.
\end{equation}
Further, the Wigner function of a single mode system corresponding to a density matrix $\hat{\rho}$ can be expressed as \citep{wigner1932quantum}:
\begin{equation}
  W(x,p)=   \frac{1}{2 \pi} \int\limits_{ -\infty}^{\infty} \bra{x+\frac{y}{2}}\hat{\rho} \ket{x-\frac{y}{2}}e^{-ipy} \hspace{0.1cm} dy.
\end{equation}
For a pure state $\ket{\psi}$ with wave function $\psi(x)$, this Wigner function simplifies to 
\begin{equation}
W(x,p)=\frac{1}{2\pi} \int\limits_{-\infty}^{\infty}  \psi(x+\frac{y}{2})\psi^{*}(x-\frac{y}{2}) e^{-ipy} \hspace{0.1cm}dy.
\end{equation}
As a consequence of the above definition, it can be seen that the Wigner function is a real-valued function. Additionally, if $W(x,p)$ is normalized, then
\begin{equation}
    \int\limits_{-\infty}^{\infty} \int\limits_{-\infty}^{\infty} W(x,p) \hspace{0.1cm} dx dp = 1.
\end{equation}

The Wigner function resembles a probability distribution but can have negative values. Therefore, it cannot be measured directly in real experiments. However, its marginals are valid probability distributions that can be realized by homodyne detection \cite{lvovsky2009continuous}. The marginal probability distributions can be obtained by integrating the Wigner function over the momentum and position space, respectively i.e., 
\begin{align}
|\psi(x)|^2 = \int\limits_{-\infty}^{\infty} W(x,p)dp,\\
    |\phi(p)|^2 = \int \limits_{-\infty}^{\infty} W(x,p)dx,
    \end{align}
where, $\phi(p)$ is the momentum wave function.

Now, to successfully develop a phase-space formalism using the Wigner function, it is necessary to establish a mapping between functions in the quantum phase-space formulation and operators in the Hilbert space within the Schr\"odinger picture. This mapping is achieved through the Weyl transform \( \tilde{A} \) of an operator \( \hat{A} \), defined as follows \cite{wigner_function_for_pedestrians}:
\begin{equation}
    \tilde{A}(x, p) = \int \limits_{-\infty}^{\infty} e^{-ipy} \langle x + \frac{y}{2} | \hat{A}(\hat{x},\hat{p}) | x - \frac{y}{2} \rangle  \hspace{0.1cm} dy.
\end{equation}
The expectation value of the operator $\hat{A}$ is,
\begin{equation}
\langle A\rangle = \int\limits_{-\infty}^{\infty}  W(x,p) \tilde{A}(x,p) \hspace{0.1cm} dx dp.
\end{equation}
Thus, the expectation value of the operator $\hat{A}$ is achieved by averaging $\tilde{A}(x,p)$ over phase-space using the quasiprobability density $W(x,p)$ that characterizes the state.

States can be classified into Gaussian and non-Gaussian based on their Wigner function. A state is considered Gaussian if its Wigner function is Gaussian, and hence positive. But the converse is not true in general. According to Hudson theorem \cite{hudson1974wigner,soto1983wigner}, pure states with positive Wigner function are always Gaussian. However, this statement does not hold true for mixed states \cite{mandilara2009extending,filip2011detecting}. There exist non-Gaussian states with a positive Wigner function,  e.g., mixture of two coherent states, mixture of Fock states \citep{walschaers_prx_quantum}. The negativity of the Wigner function is regarded as a figure of merit of non-classicality.

%%%%%%%%%%%%%%%%%%%%%%%%%%%%%%%%%%%%%%%%%%%%

\subsection{Partial transpose moments}
Entanglement is a particular type of quantum correlation that forms the basis of many information processing tasks. However, the task of entanglement detection is in general, an NP-hard problem. One of the efficient and widely used criteria for detecting bipartite entanglement in discrete variable (DV) systems is based on the positive partial transposition (PPT) criterion. This criterion examines whether the partial transposed state $\rho^{T_B}_{AB}$ (where partial transposition is taken with respect to the subsystem B) is positive semidefinite. A violation of this criterion indicates that the given state $\rho_{AB}$ is entangled. This criterion serves as a necessary and sufficient condition for ${\mathbb{C}}^2 \otimes {\mathbb{C}}^2$ and ${\mathbb{C}}^2 \otimes {\mathbb{C}}^3$ systems. However, directly computing all the eigenvalues of $\rho^{T_B}_{AB}$ is infeasible in real experiments due to computational complexity. To overcome this challenge, the concept of moments of the partially transposed density matrix (PT-moments) was introduced by Calabrese et al. in 2012 \cite{calabrese2012entanglement}. These moments provide a way to study correlations in many-body systems in relativistic quantum field theory. The eigenvalues $\{\lambda_i\}$ of $\rho^{T_B}_{AB}$ are related to the roots of its characteristic polynomial
\begin{equation}
    \text{Det}(\rho^{T_B}_{AB} - \lambda I) = \sum_{n} a_{n} {\lambda}^{n}
\end{equation}
where, $a_{n}$ are the functions of PT moments $(P_n)$ which are defined as
\begin{equation}
    P_n = \text{Tr}[{(\rho_{AB}}^{T_B})^n] \label{PTmoments}
\end{equation}
for $n=1,2,3,...$. In this way, the PT moments encapsulate information about $\{\lambda_i\}$ offering a way to express the eigenvalues in terms of experimentally accessible quantities.
In recent years, several studies have demonstrated the advantages and experimental realization of PT moments \cite{gray2018machine,carteret2005noiseless,bartkiewicz2015method,
elben2020mixed,yu2021optimal,aggarwal2024entanglement,mallick2024assessing,denis2023estimation,neven2021}. 
Note that, $P_1 = \text{Tr}[\rho_{AB}^{T_B}] = 1$. Furthermore, it can be shown that $P_2 = \text{Tr}[(\rho_{AB}^{T_B})^2]$ is related to the purity of the state. Therefore, the first non-trivial moment $P_3$ captures additional information related to the partial transposition beyond the purity of the state. Using the first three moments, a straightforward yet effective criterion for detecting entanglement has been proposed \cite{elben2020mixed}. This criterion suggests that if a state $\rho_{AB}$ is PPT, then ${P_2}^2 \leq P_3 P_1 $. Consequently, if a state violates this condition, it must be entangled, forming the 
$p_3$-PPT criterion. The $p_3$-PPT criterion and positive partial transposition (PPT) criterion are equivalent for the case of the Werner state, and hence, the $p_3$-PPT criterion is a necessary and sufficient criterion for bipartite entanglement detection of the Werner state.

%%%%%%%%%%%%%%%%%%%%%%%%%%%%%%%%%%%%%%%%%%%%%%%%%%

\section{Detection of Wigner negativity} \label{s3}
In this section, we present our main results for detecting Wigner negativity using moments of the Wigner function. We also provide explicit examples that support our proposed theorem, aiming to firmly establish its validity.\\

 \noindent\textbf{Definition 1:}  Let $W$ be the Wigner function of a $k$-mode  quantum state. We define the $m$-th order $W$-moments $(w_m) $ as:
\begin{equation} 
\begin{split}
 w_m=\int_{-\infty}^{\infty}...\int_{-\infty}^{\infty}\hspace{1mm} W ^{m} \hspace{1mm}  dx_1 \hspace{1mm} dx_2 \hspace{1mm} ...\hspace{1mm} dx_k \hspace{1mm}dp_1 \hspace{1mm} dp_2 \hspace{1mm} ...\hspace{1mm} dp_k \hspace{1mm}   \hspace{0.5cm}  \label{wmoments}
\end{split}
\end{equation}
where $m$ being an integer.

With the above definition, we are now ready to propose our criterion for detecting Wigner negativity in the following theorem.\\

\noindent\textbf{Theorem 1:} If the Wigner function $(W)$ for a $k$-mode quantum state is positive, then 
\begin{equation}
{w_2}^2 \leq w_3 \label{6}
\end{equation}
where $w_2$ and $w_3$ are defined in \eqref{wmoments}. \label{theo1}

\proof  Let the Wigner function $(W)$ be a positive, real-valued, normalized distribution of a $k$-mode  quantum state. Let us now consider the ${\mathcal{L}}_{p}$ space for $p \ge 1$, which is defined as \cite{10.5555/26851,kubrusly2007measure}
 \begin{equation}
 \begin{split}
   & ||W||_{\mathcal{L}_P} = \left(\int_{-\infty}^{\infty}...\int_{-\infty}^{\infty}\hspace{1mm} |W| ^{p} \hspace{1mm}  dx_1 \hspace{1mm} dx_2 \hspace{1mm} ...\hspace{1mm} dx_k \hspace{1mm}dp_1 \hspace{1mm} dp_2 \hspace{1mm} ...\hspace{1mm} dp_k \right)^{\frac{1}{p}} \label{schattenp}
    \end{split} 
\end{equation}
\vspace{0.1cm}

\noindent\textbf{ Hölder's inequality for ${\mathcal{L}}_{p}$ space:} Let $ 1 \leq p < \infty$ and $ 1 < q < \infty$ along with $\frac{1}{p} + \frac{1}{q} =1$. If $f \in {\mathcal{L}}_{p}$ and  $g \in {\mathcal{L}}_{q}$, then $f.g \in {\mathcal{L}}_{1}$. Moreover,
\begin{equation}
     ||fg||_{{\mathcal{L}}_1} \leq  ||f||_{{\mathcal{L}}_p} ||g||_{{\mathcal{L}}_q}.
\end{equation}
For $p=q=2$, Hölder's inequality reduces to Cauchy-Schwarz inequality.\\

In our case, we put $p=3$, $q=\frac{3}{2}$ in Hölder's inequality and taking $f=g=W$, we get
\begin{equation}
     ||W.W||_{{\mathcal{L}}_1} \leq  ||W||_{{\mathcal{L}}_3} ||W||_{{\mathcal{L}}_{\frac{3}{2}}}. \label{hoelder}
     \end{equation}
Now, 
\begin{widetext}
\begin{equation}
\begin{split}
&||W||_{{\mathcal{L}}_{\frac{3}{2}}} \\
& =\left(\int_{-\infty}^{\infty}...\int_{-\infty}^{\infty}\hspace{1mm} |W|^{\frac{3}{2}} \hspace{1mm}  dx_1 \hspace{1mm}  dx_2 \hspace{1mm} ...\hspace{1mm} dx_k \hspace{1mm}dp_1 \hspace{1mm} dp_2\hspace{1mm} ...\hspace{1mm} dp_k\right)^{\frac{2}{3}} \\
& = \left(\int_{-\infty}^{\infty}...\int_{-\infty}^{\infty}\hspace{1mm} |W|. |W|^{\frac{1}{2}}\hspace{1mm}  dx_1 \hspace{1mm} dx_2 \hspace{1mm}...\hspace{1mm} dx_k \hspace{1mm}dp_1 \hspace{1mm} dp_2 \hspace{1mm}...\hspace{1mm} dp_k\right)^{\frac{2}{3}}\\
  & \leq \left[\left(\int_{-\infty}^{\infty}...\int_{-\infty}^{\infty}\hspace{1mm} |W|^{2} \hspace{1mm}  dx_1 \hspace{1mm} dx_2 ...\hspace{1mm} dx_k \hspace{1mm}dp_1 \hspace{1mm}dp_2 ...\hspace{1mm} dp_k \right)^{\frac{1}{2}}  \left(\int_{-\infty}^{\infty}...\int_{-\infty}^{\infty}\hspace{1mm} |W| \hspace{1mm}  dx_1 \hspace{1mm} dx_2...\hspace{1mm} dx_k \hspace{1mm}dp_1 \hspace{1mm}dp_2 ...\hspace{1mm} dp_k \right)^{\frac{1}{2}}\right]^{\frac{2}{3}}  \\
   & = \left[\left(\int_{-\infty}^{\infty}...\int_{-\infty}^{\infty}\hspace{1mm} |W|^{2} \hspace{1mm}  dx_1 \hspace{1mm} dx_2 ...\hspace{1mm} dx_k \hspace{1mm}dp_1 \hspace{1mm} dp_2 ...\hspace{1mm} dp_k\right)^{\frac{1}{3}} \left(\int_{-\infty}^{\infty}...\int_{-\infty}^{\infty}\hspace{1mm} |W| \hspace{1mm}  dx_1 \hspace{1mm} dx_2 ...\hspace{1mm} dx_k \hspace{1mm}dp_1 \hspace{1mm} dp_2 ...\hspace{1mm} dp_k\right)^{\frac{1}{3}}\right]
\end{split}
\end{equation}
\end{widetext}
 %&  \hspace{10cm}\text{[using Cauchy-Schwarz inequality]} \\
i.e.,
\begin{equation}
||W||_{{\mathcal{L}}_{\frac{3}{2}}} \leq  {||W||^{\frac{2}{3}}_{{\mathcal{L}}_2}} {||W||^{\frac{1}{3}}_{{\mathcal{L}}_1}} \label{cs}
\end{equation}
Putting the value of \eqref{cs} in \eqref{hoelder}, we get,
\begin{equation}
     ||W.W||_{{\mathcal{L}}_{1}} \leq  ||W||_{{\mathcal{L}}_{3}}{||W||^{\frac{2}{3}}_{{\mathcal{L}}_2}} {||W||^{\frac{1}{3}}_{{\mathcal{L}}_1}}. \label{16}
\end{equation}
Again, 
\begin{equation}
\begin{split}
||W.W||_{{\mathcal{L}}_1} &=\int_{-\infty}^{\infty}...\int_{-\infty}^{\infty}\hspace{1mm} |W.W| \hspace{1mm}  dx_1 \hspace{1mm} ...\hspace{1mm} dx_k \hspace{1mm}dp_1 \hspace{1mm} ...\hspace{1mm} dp_k \\
& = {||W||^{2}_{{\mathcal{L}}_2}}.\\
\end{split}
\end{equation}
Therefore, it follows that
     \begin{equation}
     {||W||^2_{{\mathcal{L}}_{2}}} \leq  ||W||_{{\mathcal{L}}_{3}}{||W||^{\frac{2}{3}}_{{\mathcal{L}}_2}} ||W||^{\frac{1}{3}}_{{\mathcal{L}}_1} \label{16}
     \end{equation}
     Taking $3$rd power on both sides, we get
     \begin{equation}
     {||W||^6_{{\mathcal{L}}_{2}}} \leq  {||W||^{3}_{{\mathcal{L}}_{3}}}{||W||^{2}_{{\mathcal{L}}_2}} {||W||^{1}_{{\mathcal{L}}_1}}, \label{17}
     \end{equation}
     i.e.  \begin{equation}
     {||W||^4_{{\mathcal{L}}_{2}}} \leq  {||W||^{3}_{{\mathcal{L}}_{3}}} {||W||^{1}_{{\mathcal{L}}_1}}   \label{main}
     \end{equation}
From the normalization condition of the Wigner function $(W)$, we get 
\begin{equation}
       ||W||_{\mathcal{L}_1}  = \left(\int_{-\infty}^{\infty}...\int_{-\infty}^{\infty}\hspace{1mm} |W|^{1} \hspace{1mm}  dx_1 \hspace{1mm} ...\hspace{1mm} dx_k \hspace{1mm}dp_1 \hspace{1mm} ...\hspace{1mm} dp_k\right)^1
    =1.
\end{equation}
Since, we have assumed that the Wigner function is positive, i.e. $W = |W|$,
Therefore, from Eq.\eqref{main}, we get
\begin{equation}
     {||W||^4_{{\mathcal{L}}_{2}}} \leq  {||W||^{3}_{{\mathcal{L}}_{3}}},
     \end{equation} 
     i.e. 
     \begin{equation}
        {w_2}^2 \leq w_3,
\end{equation}
which completes our proof. \qed \\

Although our detection criteria is applicable to $k$-mode quantum systems ($k \ge 1$), for simplicity, the rest of the paper primarily focuses on single-mode and two-mode quantum systems.

Note that our proposed detection scheme is applicable to both the discrete as well as continuous variable states, provided they possess a continuous Wigner function representation. This is because the proof of Theorem 1 relies solely on the continuity properties of the Wigner function.\\

We would now like to present some examples in support of Theorem 1.

%%%%%%%%%%%%%%%%%%%%%%%%%%%%%%%%%%%%

\subsection{Examples}
\noindent\textbf{\textit{Example 1}: Two-mode squeezed vacuum state (TMSV):} This state is obtained by squeezing the two-mode vacuum state $\ket{0,0}$ using the two-mode squeezing operator $S(\zeta) = \exp(\zeta \hat{a}_{1}^{\dagger} \hat{a}_{2}^{\dagger} - {\zeta}^{*} \hat{a}_1 \hat{a}_2)$ where $\zeta = r \exp(i\phi)$ and $r > 0$ is the squeezing parameter. This state can be produced in a nondegenerate optical parametric amplifier (NOPA), and hence is represented as:
\begin{equation}
\begin{split}
    \ket{NOPA} = |\zeta\rangle = S(\zeta)\ket{0,0} = \sqrt{1-\lambda^2}\sum_{n-0}^{\infty}\lambda^n\ket{n,n} \label{nopa}
\end{split}
\end{equation} 
where, $\lambda = \tanh{r}\in [0,1]$, and $|m,n\rangle = |m\rangle \otimes |n\rangle$, where ( $|m\rangle$, $|n\rangle$ are the usual Fock states).
The Wigner function corresponding to this state is given by\cite{chowdhury2014einstein,agarwal2012quantum,maity2017tighter},
\begin{equation}
\begin{split}
     W_{\zeta}[x_1,p_1,x_2,p_2]  =&  \frac{1}{\pi^2} \text{exp}[ - 2\sinh{[2r]}(x_1 x_2-p_1 p_2) \\& - \cosh{[2r]}({x_1}^2+{x_2}^2+{p_1}^2+{p_2}^2)].
\end{split}
\end{equation}
The Wigner function of the two-mode squeezed vacuum state is always positive \cite{agarwal2012quantum}. Therefore, from Theorem 1, we should have $\omega_2^2 \le \omega_3$.
Now, we calculate the second $(\omega_2)$ and third order moments $(\omega_3)$ of the Wigner function as follows:
\begin{equation}
\begin{split}
\omega_2^2 & =\left(\int\limits_{ -\infty}^{\infty} \int\limits_{ -\infty}^{\infty}\int\limits_{-\infty}^{\infty}\int\limits_{-\infty}^{\infty} W_{\zeta}^{2} \hspace{0.15cm} {dx}_1 dp_{1}  {dx}_2  dp_{2}\right)^{2} 
\\& = \frac{1}{16 \pi^4}.
\end{split}
\end{equation}
\begin{equation}
\begin{split}
\omega_3 & =\left(\int\limits_{ -\infty}^{\infty} \int\limits_{ -\infty}^{\infty}\int\limits_{-\infty}^{\infty}\int\limits_{-\infty}^{\infty} W_{\zeta}^{3} \hspace{0.15cm} {dx}_1 dp_{1}  {dx}_2  dp_{2}\right) 
\\&= \frac{1}{9 \pi^4}
\end{split}
\end{equation}
Clearly,  $\omega_2^2 \le \omega_3$, which validates our proposed criterion.
\vspace{0.2cm}

\noindent\textbf{\textit{Example 2}: Single photon-subtracted squeezed vacuum (SPSSV):} Let us now consider an explicit example of a non-Gaussian state that is obtained by subtracting photons from a Gaussian state. The subtraction of $n$ photons from the state $|\zeta\rangle$ (Eq. \eqref{nopa}) may be represented as,
\begin{equation}
    |\zeta_{-n}\rangle = [a \otimes I + (-1)^k I \otimes b]^n|\zeta\rangle
\end{equation} 
where $k\in \{0,1\}$. It is also assumed that the specific mode from which the photon is subtracted remains unknown. SPSSV state is generated by subtracting a single photon from any mode of a two-mode squeezed vacuum state, represented as \cite{biswas2007nonclassicality}, 
\begin{equation}
\begin{split}
    \ket{\xi_{-1}}= & \frac{1}{2\sinh{r}^2}\sqrt{1-\lambda^2}\sum_{n=0}^{\infty}\lambda^n\sqrt{n} [ \ket{n-1,n} \\ & +(-1)^k\ket{n,n-1} ]
    \end{split}
\end{equation}
with $k\in [0,1]$ and $\lambda = \tanh{r}, \hspace{0.1cm} r >0$.

The Wigner function of SPSSV state \cite{agarwal2012quantum}, can be represented as
\begin{equation}
\begin{split}
   W_{1}[x_1,p_1,x_2,p_2] =& \frac{1}{\pi^2} \text{exp}[2\sinh{[2r]}(x_1 x_2-p_1 p_2) \\& - \cosh{[2r]}({x_1}^2+{x_2}^2+{p_1}^2+{p_2}^2)] \nonumber \\
    &\times [-\sinh{[2r]}((p_1-p_2)^2-(x_1-x_2)^2) \\ &+ \cosh{[2r]}((p_1-p_2)^2+(x_1-y_1)^2)-1].
\end{split}
\end{equation}
 It is well known that the state exhibits Wigner negativity \cite{agarwal2012quantum}, therefore from our proposed criteria we should have $\omega_2^2 > \omega_3$. Now, we compute the second $(\omega_2)$ and third order moments $(\omega_3)$ of the Wigner function as follows:
 % taking $r=1$, let us calculate $\omega_2^2$ and $\omega_3$,
\begin{equation}
\begin{split}
\omega_2^2 & =\left(\int\limits_{ -\infty}^{\infty} \int\limits_{ -\infty}^{\infty}\int\limits_{-\infty}^{\infty}\int\limits_{-\infty}^{\infty} W_{1}^{2} \hspace{0.15cm} {dx}_1 dp_{1}  {dx}_2  dp_{2}\right)^{2} 
\\&= \frac{1}{16 \pi^4}
\end{split}
\end{equation}
\begin{equation}
\begin{split}
\omega_3 &=\left(\int\limits_{ -\infty}^{\infty} \int\limits_{ -\infty}^{\infty}\int\limits_{-\infty}^{\infty}\int\limits_{-\infty}^{\infty} W_{1}^{3} \hspace{0.15cm} {dx}_1 dp_{1}  {dx}_2  dp_{2}\right)
\\&= \frac{1}{81 \pi^4}
\end{split}
\end{equation}
Clearly, $\omega_2^2 > \omega_3$ which again supports Theorem 1.

\vspace{0.2cm}

\noindent\textbf{\textit{Example 3}: NOON State:} This is an example of a two-mode state where $N$ photons can be found either in the mode $a$ or in the mode $b$. The NOON state is %utilized in making precise interferometric measurements in quantum metrology. 
a maximally path entangled number state with the form
\begin{equation}
    \begin{split}
        |\psi\rangle = \frac{1}{\sqrt{2}}(|N\rangle_{a} |0\rangle_{b} + e^{i\phi}|0\rangle_{a} |N\rangle_{b}).
    \end{split}
\end{equation}
The Wigner function for this state with $\phi = \pi$, in terms of dimensionless quadratures ${x_1, p_1}$ and ${x_2, p_2}$ is given by \cite{agarwal2012quantum},
\begin{equation}
    \begin{split}
         W_{N}[x_1,p_1,x_2,p_2] =& \frac{1}{2\pi^{2}N!} e^{-({x_1}^2+{x_2}^2+{p_1}^2+{p_2}^2)}\\& \times [ -2^{N} \{ (x_1 + ip_1)^{N} (x_2 - ip_{2})^{N} \\& +  (x_1 - ip_1)^{N} (x_2 + ip_{2})^{N} \} \\& +(-1)^{N} N!\{ L_{N}[2({x_1}^2+{p_1}^2)] \\& + L_{N}[2({x_2}^2+{p_2}^2)]\}].
    \end{split}
\end{equation}
Here $L_N (x)$ is the Laguerre polynomial. The Wigner function of the NOON state is a well-defined quasi-probability distribution function, within the infinite-dimensional continuous phase space of position $(x)$ and momentum $(p)$. 
This continuity holds despite the inherent discrete nature of the NOON state itself. Since our moment-based criterion, as proposed in Theorem 1, is entirely based on the Wigner function of an unknown quantum state, it is also applicable for detecting Wigner negativity in  discrete-variable states such as the NOON state which possess a continuous Wigner function representation.

\begin{figure}
    \centering
    \includegraphics[width=8cm]{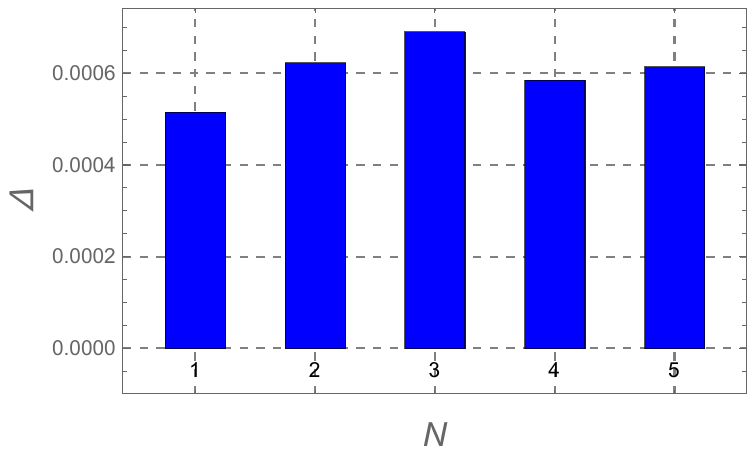}
    \caption{Plot of $\Delta$ = $ w_2^2 - w_3 $ with the photon number $(N)$ for NOON states upto $N=5$.}
    \label{fig:1}
\end{figure}

%{\color{red} The Wigner function of NOON state is a well-defined quasi-probability distribution function, within the infinite-dimensional continuous phase space of position $(x)$ and momentum $(p)$. This continuity holds despite the inherent discrete nature of the NOON state itself. Hence the same condition stated in Theorem 1 holds true even for discrete-variable states that have a continuous Wigner function representation.\\
Now, this state exhibits Wigner negativity for any $N\geq 1$. Therefore, for $N\geq 1$, we should have 
${w_2}^2 - w_3 >0$ which is evident from Fig. \ref{fig:1}, where $w_2$ and $w_3$ are defined in \eqref{wmoments}.
The explicit values of the moments of Wigner function and $w_{2}^{2}-w_{3}$ are presented in table \ref{table1} of appendix \ref{A}.
\vspace{0.2cm}

\noindent\textbf{\textit{Example 4}: Fock States:} Another important class of non-Gaussian states is the Fock states. Let $|n\rangle \equiv {\frac {a^{\dagger n}}{\sqrt {n!}}}|0\rangle $ be the ${n}$-th Fock state of a quantum harmonic oscillator. The associated Wigner function in dimensionless variables is given by \cite{agarwal2012quantum},
\begin{equation}
    \begin{split}
        {W_{|n\rangle }(x,p)={\frac {(-1)^{n}}{\pi }}e^{-(x^{2}+p^{2})}L_{n}{\big (}2(p^{2}+x^{2}){\big )}}
    \end{split}
\end{equation}
where ${ L_{n}(x)}$ denotes the ${n}$-th Laguerre polynomial.

\begin{figure}
    \centering
    \includegraphics[width=8cm]{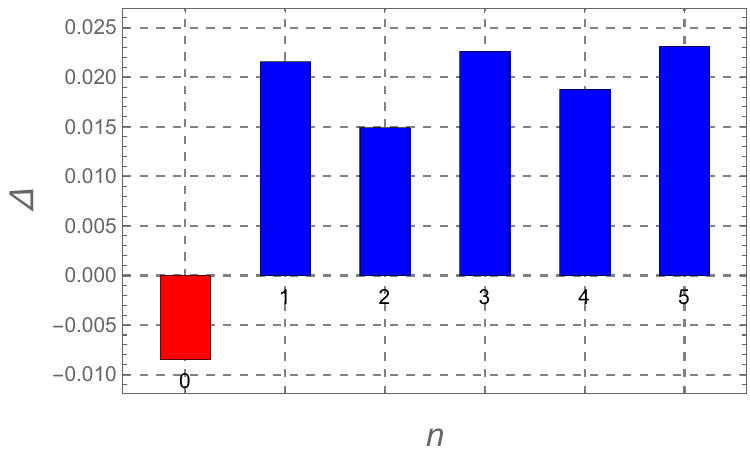}
    \caption{$\Delta$ = $ w_2^2 - w_3 $ is plotted versus the number of the Fock state up to $n=5$. The figure clearly shows Wigner negativity (solid blue) for all states except $n=0$ (solid red). }
    \label{fig:2}
\end{figure}

The vacuum state $|0\rangle$ is a Gaussian state, hence it possesses a positive Wigner function. All other Fock states are Wigner negative. Therefore, for $n\geq 1$, we should have 
${w_2}^2 - w_3 >0$ which is again evident from Fig. \ref{fig:2}.
The moments of the Wigner function for the first five Fock states are provided in table \ref{table2} of appendix \ref{A}.\\

It may be interesting to note that there exist certain mixed Wigner-negative states that can not be detected using our criterion proposed in Theorem 1. For example, consider the mixture of Fock states with $n=0$ (which has a positive Wigner function) and $n=1$ (which has a negative Wigner function) as following:
\begin{align}
    \rho = \lambda \ket{0}\bra{0} +(1-\lambda)\ket{1}\bra{1}.
\end{align}
The corresponding Wigner function for this state \cite{walschaers_prx_quantum}
\begin{align}
    W(x,p) = \frac{1}{2\pi} [ (1-\lambda) (x^2 +p^2) +2\lambda - 1] e^{-\frac{1}{2} (x^2 +p^2)}
\end{align}
turns out to be negative for $0<\lambda<0.5$. However, we obtain $\omega_2^2 > \omega_3$ for the region $0<\lambda \le 0.31$, and therefore our criterion fails to detect negativity of the Wigner function for this state in the range $0.31<\lambda<0.5$.

%%%%%%%%%%%%%%%%%%%%%%%%%%%%%%%%%%%%%%%%%%%%%%
\section{Experimental proposal for calculating moments of the Wigner function}\label{s4} 

 In this section, we propose a method for experimentally measuring the moments of the Wigner function. To determine the $m$-th order moments of the Wigner function, $m$ copies of the unknown quantum state are required. Our approach involves using the SWAP operator between the individual constituents of these $m$ copies i.e. Alice (Alice and Bob)  apply the SWAP operator locally within her (their) (sub)system(s) for calculating the moments corresponding to a single mode (two-mode) quantum state. The mode SWAP operator can be written as \cite{volkoff2022ancilla}
\begin{equation} \label{SWAP}
    \text{SWAP}= \text{exp}\left[{{i \pi}{(\frac{{\hat{a}^{\dagger}_1} - {\hat{a}^{\dagger}_2}}{\sqrt{2}}) (\frac{{\hat{a}_1} - {\hat{a}_2}}{\sqrt{2}})}}\right]
\end{equation}
where, $\hat{a}_k$ and ${\hat{a}^{\dagger}_k}$ are the $k$-th mode annihilation and creation operators, for $k =1,2$. The annihilation and creation operators in the canonical basis can be written as
\begin{equation} \label{creation}
    \begin{split}
        \hat{a}_k = \frac{(\hat{x}_k + i \hat{p}_k)}{\sqrt{2}} \hspace{2mm},\hspace{2mm}{\hat{a}_k }^{\dagger} = \frac{(\hat{x}_k - i \hat{p}_k)}{\sqrt{2}}
    \end{split}
\end{equation}
with $[ \hat{x}_k, \hat{p}_k ] = i \mathbb{I}$. The action of the SWAP operator is 
     $\text{SWAP}a_{1(2)}^\dagger\text{SWAP}^\dagger = a_{2(1)}^\dagger$.
% \end{equation}
This operator is not tied to the discrete or continuous nature of the state but rather operates on the mode labels themselves. The mode SWAP operator works equivalently for both CV and DV states because it acts on the mode indices.
Using Eqs.\eqref{SWAP} and \eqref{creation}, we get
\begin{equation}
    \text{SWAP}= \text{exp}\left[{\frac{i \pi}{4}}{\left({{(\hat{x}_1-\hat{x}_2)^2 + (\hat{p}_1-\hat{p}_2)^2 -2\mathbb{I} }}\right)}\right].
\end{equation}
Now, we can do the Weyl transform of this mode SWAP operator to get the SWAP function $(f_{\text{SWAP}})$,
\begin{equation}
\begin{split}
 f_{\text{SWAP}} &= \int\limits_{-\infty}^{\infty} \int\limits_{-\infty}^{\infty} dy_1 dy_2 \bigg [ e^{-i(p_1y_1 + p_2y_2)} \\
  &\times \left\langle x_1 + \frac{y_1}{2},   x_2 + \frac{y_2}{2}\left| {\text{SWAP}} \right| x_1 - \frac{y_1}{2},x_2 - \frac{y_2}{2} \right\rangle \bigg].
\end{split}
\end{equation}

\noindent\textbf{Realization of the second order moment $(w_2)$:}  The second order moment of the Wigner function of an unknown two-mode quantum state $\ket{{\psi}}_{AB}$ is defined as
\begin{equation}
     w_2=\int\limits_{-\infty}^{\infty} \int\limits_{-\infty}^{\infty}\int\limits_{-\infty}^{\infty}\int\limits_{-\infty}^{\infty} W^{2} \hspace{0.1cm} {dx}_1 {dp}_1 dx_{2} dp_{2}.
\end{equation}
To calculate the second order moment $(w_2)$, we take two copies of $\ket{{\psi}}_{AB}$. 
 Let $C_2= A_1 B_1  A_2 B_2 $ be the CV registrar acting on the two-mode registers $A$ and $B$ consisting of isomorphic modes $A_1,A_2$ and $B_1,B_2$ respectively. The first step of the protocol is to express $w_2$ as an expectation value of the SWAP operator, similar to the procedure in \cite{gray2018machine}, i.e.,
\begin{equation}
 w_2 = \int\limits_{-\infty}^{\infty} \int\limits_{-\infty}^{\infty}\int\limits_{-\infty}^{\infty}\int\limits_{-\infty}^{\infty} (W^{\otimes2}) \hspace{.1cm}   f_{\text{SWAP}} \hspace{0.2cm} {dx}_1 {dp}_1 dx_{2} dp_{2}
 \end{equation}
 Such a  method of expressing the second order moment has been used extensively in entanglement theory, where a SWAP operator has been employed to realize this in a simple quantum network \cite{ekert2002direct}. Linear and non-linear functionals of any density operator can be directly estimated by this procedure \cite{horodecki2002method}. This has been extended in the domain of continuous systems in a similar fashion following the prescription of \cite{volkoff2022ancilla} and \cite{hastings2010measuring}.

Now, for the case of a single mode quantum state, Alice applies the mode SWAP operator on her two particles to obtain the expectation value. 
In case of a two mode state, Alice and Bob locally apply the SWAP operator on their respective subsystems of $\ket{{\psi}}_{AB}^{\otimes 2}$, i.e., Alice and Bob each apply the mode SWAP operator on their two particles to obtain the expectation value.\\

\noindent\textbf{Realization of the third order moment $(w_3)$:}  The third order moment of the Wigner function of an unknown quantum state $\ket{{\psi}}_{AB}$ is defined as 
\begin{equation}
     w_3=\int\limits_{-\infty}^{\infty} \int\limits_{-\infty}^{\infty}\int\limits_{-\infty}^{\infty}\int\limits_{-\infty}^{\infty} W^{3} \hspace{0.1cm} {dx}_1 {dp}_1 dx_{2} dp_{2}.
\end{equation}
The third order moment is calculated using three copies of  $\ket{{\psi}}_{AB}$ and by implementing the CV registrar $(C_3= A_1 B_1  A_2 B_2 A_3 B_3) $ on the two-mode registers A and B consisting of isomorphic modes $A_1, A_2, A_3$ and $B_1, B_2, B_3$ respectively. Let, ${\mathbb{P}}^{(3)} = \prod _{i=1}^3 \text{SWAP}_{A_{i},A_{i+1}} (\prod _{i=1}^3 \text{SWAP}_{B_{i},B_{i+1}})$ be the CV permutation operator acting on the register $A$ ($B$) (where $3 + 1$ is taken modulo $3$). We express $w_3$ as the expectation value of the permutation operator as follows \cite{volkoff2022ancilla, hastings2010measuring, gray2018machine}
\begin{equation}
     w_3 =
       \int\limits_{-\infty}^{\infty} \int\limits_{-\infty}^{\infty}\int\limits_{-\infty}^{\infty}\int\limits_{-\infty}^{\infty} (W^{\otimes3}) \hspace{.1cm}  {f_{\mathbb{P}^{(3)}}} \hspace{0.2cm} {dx}_1 {dp}_1 dx_{2} dp_{2} 
\end{equation}
where ${f_{\mathbb{P}^{(3)}}} = \prod _{i=1}^3 f_{\text{SWAP}_{A_{i},A_{i+1}}} (\prod _{i=1}^3 f_{\text{SWAP}_{B_{i},B_{i+1}}})$ is the function corresponding to CV permutation operator 
acting on the register $A (B)$. 

 In the case of a single-mode quantum state, three copies of the unknown quantum state are required. The SWAP operator, $\text{SWAP}_{A_{i}, A_{i+1}}$ (with $i\in \{1,2\}$) can be applied in a forward sequence on Alice's particles between neighboring copies to obtain the expectation value.

In case of two-mode systems, measuring $w_3$ necessarily involves the following steps :
\begin{itemize}
    \item Prepare three copies of $\ket{{\psi}}_{AB}$.
    \item Apply SWAP operator in a forward sequence on Alice's side $\text{SWAP}_{A_{i}, A_{i+1}}$  between neighboring copies for $i=1,2$.
    \item Apply adjacent SWAP operator in a backward sequence on Bob's side $\text{SWAP}_{B_{i}, B_{i+1}}$ between neighboring copies, ranging from $b=3$ to $b=2$.
    \item Follow these steps to obtain the expectation value.
\end{itemize} 
{\color{red}}

Computation of these moments do not require prior knowledge of the state, and are experimentally less demanding compared to existing methods such as tomography.
%%%%%%%%%%%%%%%%%%%%%%%%%%%%%%%%%%%%%%%%%%%%%%%
\vspace{0.5cm}
\section{Conclusions}\label{s5}

Non-classical states that exhibit properties beyond classical physics, are essential for various quantum information processing tasks \cite{niset2009no,eisert2002distilling,bartlett2002efficient,galvao2005discrete,shang2024resonance}. However, before utilizing them for any information processing protocol, it is crucial to detect the signature of the specific type of non-classicality pertaining to each such state. In this paper, we have focused on detecting a particular type of such non-classicality known as Wigner negativity. We
have shown how Wigner negativity could be detected through evaluating simple functionals using the first
three moments of the Wigner function.
We have illustrated our approach through various examples of well-known 
single-mode and two-mode discrete variable as well as
continuous variable states.

 Our results demonstrate that calculating only the first three moments of the Wigner function is sufficient to identify such non-classical traits reliably. Unlike tomography or witness-based approaches, our method does not require 
 prior knowledge of the state or reconstruction of the Wigner function \cite{chabaud2021witnessing,lvovsky2009continuous,d2001quantum, d2003quantum}. Instead, it relies on classical shadows of the state. We have further proposed an experimental method for calculating these moments based
 on application of the SWAP operation. Our proposal suggests that these moments can be computed efficiently in real experiments, establishing the protocol's experimental feasibility.

Our present study motivates several  directions for future analysis. Specifically, akin to entanglement detection \cite{gray2018machine,elben2020mixed,neven2021,yu2021optimal}, a higher order moment-based detection scheme could prove highly relevant for identifying Wigner negativity in mixed states for which moments up to the third order
are insufficient. Exploring and developing a detailed derivation and analysis of the moment-based  scheme for multi-qubit or high-dimensional optical states would be an important extension of our current study, enabling the
detection of Wigner negativity of such states with exponentially fewer samples than standard tomography. Moreover, extending such moment-based detection schemes to other types of non-classicality is also significant. Utilization of such schemes for continuous variable (CV) states seems promising in effectively characterizing  devices, particularly those consisting of a large number of particles and optical setups.

\section{Acknowledgements}\label{s6}
B.M. acknowledges the DST INSPIRE fellowship program for financial support.

%%%%%%%%%%%%%%%%%%%%%%%%%%%%%%%%%%%%%%%%%%%%%%%
\bibliography{wignernegativity}
\appendix
\section{Calculation of moments of Wigner function for NOON and Fock states} \label{A}
Here, we present two tables demonstrating the effectiveness of our detection criteria for detecting Wigner negativity in NOON states and Fock states respectively. The tables provide explicit values of the second and third order moments of the Wigner function ($w_2$ and $w_3$ respectively) for different numbers of photons that are used for Fig. \ref{fig:1} and Fig. \ref{fig:2} respectively.
\newpage
\begin{table}
\begin{tabular}{|>{\centering\arraybackslash} p{0.5cm}|>{\centering\arraybackslash}p{2.5cm}|>{\centering\arraybackslash}p{2.5cm}|>{\centering\arraybackslash}p{2.5cm}|} \hline 
 \vspace{0.02cm}
 $N$ &  $ w_2 $ & $ w_3 (\times 10^{-4} )$ & $ w_2^2 - w_3 ( \times 10^{-4})$ \\ \hline 
 \vspace{0.02cm}
  1   &   $ 0.025330 $    &   $1.26741$    &    $5.14883$ \\ \hline 
 \vspace{0.02cm}
 2   &   $ 0.025330 $    &    $0.140823$    &    $6.27542$ \\ \hline 
 \vspace{0.02cm}
 3   &   $ 0.025330 $     &    $ - 0.488533 $    &    $ 6.90477 $ \\ \hline 
 \vspace{0.02cm}
 4   &   $ 0.025330 $     &   $0.592074$    &    $5.82416$ \\ \hline
  \vspace{0.02cm}
 5   &   $ 0.025330 $     &   $0.271104$    &    $6.14693$ \\ \hline
 
\end{tabular} \\
\caption{\label{tab:table-name}  Computation of $w_2$, $w_3$ and $w_{2}^{2}-w_3$ for the number of photons, $N \in \{1,2,3,4,5\}$ for NOON states.}\label{table1}
\end{table}
\begin{itemize}
\item {\bf NOON states:}
From the table, it is evident that the values of $w_2^2 - w_3$ are always positive for the five NOON states, all of which possess Wigner negativity.
\end{itemize}
\begin{table}
\begin{tabular}{|>{\centering\arraybackslash} p{0.5cm}|>{\centering\arraybackslash}p{2.5cm}|>{\centering\arraybackslash}p{2.5cm}|>{\centering\arraybackslash}p{2.5cm}|} \hline
\vspace{0.02cm}
$n$ & $ w_2 $ & $ w_3 $ & $ w_2^2 - w_3 $ \\
 \hline
 \vspace{0.02cm}
 $0$   &   $0.159155$&   $0.033774$&    $-0.008443$ \\ \hline 
\vspace{0.02cm}
 $1$   &   $0.159155$&   $0.003753$&    $0.021578$ \\ \hline 
\vspace{0.02cm}
 $2$   &   $0.159155$&    $0.010424$&    $0.014906$ \\ \hline 
 \vspace{0.02cm}
 $3$   &   $0.159155$&    $0.002723$&    $0.022607$ \\ \hline
\vspace{0.02cm}
 $4$   &   $0.159155$&   $0.006548$&    $0.018782$ \\ \hline
 \vspace{0.02cm}
$5$   &   $0.159155$&    $0.002222$&    $0.023108$ \\ \hline

\end{tabular} \\
\caption{\label{tab:table-name} Computation of $w_2$, $w_3$ and $w_{2}^{2}-w_3$ for the photon number, $n \in \{0,1,2,3,4,5\}$ for Fock states.}\label{table2}
\end{table}
\begin{itemize}
\item {\bf Fock states:}
The table reveals that the value of $w_2^2 - w_3$ is negative for $n=0$, which is a Gaussian state (positive Wigner function), and positive for other Fock states which exhibit Wigner negativity.

Hence, these results are in agreement with our detection criterion.
\end{itemize}

\end{document}